\documentclass[10pt, conference, letterpaper]{IEEEtran}

\usepackage{comment}
\usepackage[T1]{fontenc}
\usepackage[utf8]{inputenc}
\usepackage{tensor}
\usepackage[cmex10]{amsmath} 
\usepackage{physics}
\usepackage{amssymb,amsfonts}
\usepackage{mathtools, physics, bbm}

\usepackage{amsthm}

\newtheorem{definition}{Definition}

\newtheorem*{Game*}{Game}
\newtheorem*{TokGen*}{TokenGeneration Phase}
\usepackage{breqn}
\newtheorem*{TokVer*}{TokenVerification Phase}
\newtheorem*{Setup*}{Setup}
\newtheorem*{Query*}{Query}
\newtheorem*{Challenge*}{Challenge}
\newtheorem*{Guess*}{Guess}
\usepackage{breqn}
\theoremstyle{remark}

\usepackage{tikz}
\usepackage{tabularx}
\usetikzlibrary{quantikz}
\usepackage{algorithm}
\usepackage{algpseudocode}
\usetikzlibrary{quantikz}

\newcounter{protocol}
\newenvironment{protocol}[1]
  {\par \addvspace{\topsep}
   \noindent
   \tabularx{\linewidth}{@{} X @{}} \\
    \hline
    \stepcounter{protocol}
    \textbf{protocol \theprotocol } #1 \\
    \hline}
  { \\
    \hline
   \endtabularx
   \par\addvspace{\topsep}}

\usepackage{tikz}
\usetikzlibrary{quantikz}

\usepackage{graphicx}
\usepackage{url}
\usepackage{cite}
\usepackage{hyperref}
\hypersetup{
    hidelinks = true
}
\usepackage[capitalize]{cleveref}

\usepackage{algorithm}
\usepackage{algpseudocode}
\usepackage{amssymb}

\begin{document}

\title{
Feasibility of Logical Bell State Generation in Memory Assisted Quantum Networks}

\author{
\IEEEauthorblockN{Vladlen Galetsky}
\IEEEauthorblockA{TUM, Germany\\
vladlen.galetsky@tum.de} 
\and
\IEEEauthorblockN{Nilesh Vyas}
\IEEEauthorblockA{Airbus Central R\&T, Germany\\
 nilesh.vyas@airbus.com}
\and
\IEEEauthorblockN{Alberto Comin}
\IEEEauthorblockA{Airbus Central R\&T, Germany\\
alberto.comin@airbus.com}
\and
\IEEEauthorblockN{Janis Nötzel}
\IEEEauthorblockA{TUM, Germany\\
janis.noetzel@tum.de} 
}
\maketitle

\begin{abstract}
This study explores the feasibility of utilizing quantum error correction (QEC) to generate and store logical Bell states in heralded quantum entanglement protocols, crucial for quantum repeater networks. Two lattice surgery-based protocols (local and non-local) are introduced to establish logical Bell states between distant nodes using an intermediary node. We simulate the protocols using realistic experimental parameters, including ion trap memories, noisy optical channels, frequency conversion and non-destructive detection of photonic qubits. The study evaluates rotated and planar surface codes alongside Bacon-Shor codes for small code distances (\(d = 3, 5\)) under depolarizing and physical noise models.  Pseudo-thresholds are identified, with physical error rates above  \(p_{\text{err}} \sim 10^{-3}\) offering no advantage over unencoded Bell states under depolarizing noise. Pseudo-thresholds are also reevaluated in terms of gate error rates \(p_{\text{err}_H}\), \(p_{\text{err}_{CX}}\) and \(p_{\text{err}_M}\). For a distance of 1 km between the end node and the intermediary, an advantage over unencoded Bell-state heralded protocols requires reducing gate error rates by an order of magnitude (\(0.1p_{\text{err}_H}\), \(0.1p_{\text{err}_{CX}}\), and \(0.1p_{\text{err}_M}\)). These results highlight the need for significant hardware improvements to implement logical Bell state protocols with quantum memories. 
Additionally, the non-local protocol rate was analyzed achieving rates up to \((32.53\pm1.53) \, \mathrm{Hz}\) over distances of \(1\) to \(80\) \(\mathrm{km}\) between the end node and the intermediary node.



\end{abstract}

\begin{IEEEkeywords}
Quantum repeaters, quantum networks, lattice surgery, logical Bell-pairs, logical heralded entanglement protocol, Bacon-Shor codes, surface codes.
\end{IEEEkeywords}

\section{Introduction}
The vision for a quantum internet, as outlined in \cite{QuantumInternet,ILLIANO2022109092}, is fundamentally rooted in fault-tolerant quantum communication. This requires the integration of quantum repeaters equipped with highly efficient and robust quantum memories for terrestrial links, and free-space quantum links facilitated by satellite-ground communication. Quantum repeater architectures have been meticulously designed \cite{Muralidharan2015OptimalAF, AzumaQR} to address errors primarily arising from photon loss, where photons are either absorbed or scattered and gate operation errors, which result from device imperfections causing noise and reduced fidelity.

Logical Bell-pair generation is a fundamental building block for quantum repeater networks, ensuring high fidelity and reliable entanglement distribution during transmission and processing. Consequently, quantum repeater architectures are classified into three generations, each introducing progressively advanced error correction methods \cite{Muralidharan2015OptimalAF, AzumaQR}.  Quantum repeaters, utilizing probabilistic error suppression to manage practical imperfections \cite{Azuma_2012}, have seen significant progress in recent times. Milestone experiments have demonstrated heralded entanglement distribution between two absorptive quantum memories \cite{Liu2021-jh, Lago_Rivera_2021}, realization of a multimode quantum network of remote solid-state qubits \cite{Pompili_2021}, entanglement of trapped-ion qubits separated by $230$ $\mathrm{m}$ \cite{Krutyanskiy_2023}. However, these demonstrations are limited in achieving the high fidelity and long-distance communication essential for a scalable quantum internet \cite{QuantumInternet, AzumaQR, ILLIANO2022109092}.

Incorporating quantum error correction (QEC) to mitigate errors signifies a major leap forward, promising more robust and scalable quantum networks. However, despite their immense potential, there has been relatively limited research on the implementation of QEC for generating and storing logical Bell-pairs in a quantum network scenario, compared to the extensive experimental efforts on heralded entanglement generation. This highlights the critical need for a comprehensive feasibility study on the generation of logical Bell-pairs and their storage in quantum memories.

This work endeavors to bridge the existing research gap by conducting detailed simulations on generating logical Bell-pairs using lattice surgery within a noise and error model that incorporates experimental parameters. We design two new memory-dependent logical heralded entanglement protocols relying on a local and non-local logical Bell-pair generation, respectively and we study their performance while evaluating the practical challenges associated with various QEC codes. We aim to gain vital insights into their viability and identify key experimental parameters requiring further development.

\section{Related Work}
\label{relatedwork}

Intensive research and efforts are dedicated to enhancing quantum hardware and protocols for increased robustness and reliability. Recent advancements in quantum memory platforms have been pivotal in enabling various quantum communication protocols.  For instance, the work by \cite{PRXQuantum101km} has demonstrated the feasibility of long-lived quantum memory, enabling single 87Rb atom-photon entanglement over $101$ $\mathrm{km}$ of telecom fiber, marking a significant step forward in neutral atom quantum memories. In another notable study, \cite{Langenfeld} developed a quantum repeater node with two memory atoms in an optical cavity, highlighting the potential of memory atoms in enhancing the security of quantum communication networks. The work of \cite{liu2023} successfully demonstrated heralded entanglement generation between two quantum nodes situated $12.5$ $\mathrm{km}$ apart, with the entanglement storage time exceeding the round-trip communication time. Additionally, \cite{kucera2024} reported on the implementation of quantum entanglement distribution and quantum state teleportation over a $14.4$ $\mathrm{km}$ urban dark-fiber link. This work exemplifies the application of quantum teleportation in real-world urban environments, bridging the gap between theoretical proposals and practical implementations.

The experimental application of entanglement-based protocols using topological quantum error correction (QEC) codes is a relatively new and evolving field. Notably, Erhard et al. \cite{Erhard2021} demonstrated local logical quantum teleportation using lattice surgery with two \(d=2\) surface codes on an ion trap quantum processor, highlighting the potential for applying such methods in non-local quantum network scenarios if further investigated. Similarly, Luo et al.  \cite{Luo_2021} showcased logical quantum teleportation, leveraging entangled offline resource states to execute non-transverse gates. More recently \cite{bsc}, hybrid surface and Bacon-Shor codes have been explored for creating local logical Bell states, as studied in the context of the Heavy-Hex Lattice architecture, with demonstrations on IBM's 133-qubit Torino processor.

From an information-theoretic perspective, monolithic surface code quantum communication was first introduced by Fowler et al.\cite{PhysRevLett.104.180503}, although without a realistic error model or memory considerations. Subsequently, surface code applications in quantum networks were analyzed by Vuillot et al. \cite{surfacenet}, employing graph-like approaches but without entangled states.

The advancement of QEC and quantum networks has been supported by various simulation tools.
 For QEC, simulators like Stim \cite{Gidney2021stimfaststabilizer} and some high performance compilers \cite{Watkins_2024, Paler2016} allow to accurately simulate quantum stabilizer circuits. For quantum network simulations, tools like NetsQuid \cite{Coopmans2021}, SeQUeNCe \cite{SeQUeNCe}, QuReed
\cite{sekavčnik2024qureed}, QuNetSim \cite{QuNetSim}, SQUANCH \cite{SQUANCH}, and SimulaQron \cite{SimulaQron} facilitate virtual testing and optimization of quantum network protocols.

\section{Our contributions}
 
This study expands upon the latest developments in heralded quantum entanglement generation, a building block of any quantum repeater protocol. By utilizing QEC codes in quantum memories, we correct logical Bell states in a fault-tolerant manner. We analyze the feasibility, challenges, and benefits of generating and storing these states through detailed simulations incorporating realistic experimental parameters. This work is the first to analyze the practical feasibility of non-local logical Bell states in quantum networks, with quantum error correction (QEC) performed in a quantum memory. 

We study a configuration where Alice and Bob, two distant nodes, aim to establish a logical Bell-pair with the help of an intermediary node, Charlie. This work introduces two protocols for executing the heralded entanglement protocol with logical Bell-pairs: a local and a non-local scheme. Both protocols use heralding techniques to address loss errors and employ quantum error correction via lattice surgery to construct and correct a logical Bell state. In the local protocol, Charlie performs lattice surgery to generate the logical Bell state locally and directly transmits it to Alice and Bob, storing the Bell states in quantum memory for later measurement. In the non-local protocol, Charlie distributes physical entanglement pairs to Alice and Bob, who independently perform lattice surgery during $d$ merging cycles to generate a logical Bell state in a distributed manner.

Our simulations incorporate ion trap quantum memories, noisy optical channels, quantum non-demolition measurements, frequency conversion, photon capture and state transfer accounting for various experimental parameters (Table\,\ref{parameter:table}). We evaluate the feasibility of generating and storing logical Bell states using $d=3$ and $d=5$ surface and Bacon-Shor codes within this simulation framework.

We consider the depolarization channel, Section\,\ref{standard_noise_channel}, to identify pseudo-thresholds where physical error rates exceeding $p_{\text{err}} = (5.5\pm 0.2)  \times 10^{-4}$ for $S[[18,2,3]]$, $p_{\text{err}} = (9.0\pm 0.3)  \times 10^{-4}$ for rotated $S[[18,2,3]]$, and $p_{\text{err}} = (1.5\pm 0.2) \times 10^{-3}$ for the $BS[[18,2,3]]$ codes, provide no advantage over unencoded Bell states in either protocol. These findings extend the results of \cite{PhysRevA.98.050301}, which reported pseudo-thresholds of $p_{\text{err}} = 9.0\times10^{-3}$ for Bacon-Shor codes and $p_{\text{err}} = 1.5\times10^{-3}$ for surface codes, both assessed without lattice surgery and idle qubit noise consideration. We also determined the thresholds at which increasing the code distance worsens the logical error rate (\(p_L\)). By comparing \(d=3\) with \(d=5\) codes, we found thresholds of \(p_{\text{err}} = (3.9 \pm 0.1) \times 10^{-3}\) for \(BS\) codes and \(p_{\text{err}} = (5.8 \pm 0.2) \times 10^{-3}\) for rotated \(S\) codes.

Using a more realistic noise model, as described in Section\,\ref{robustn}, we reevaluated these thresholds in terms of $H$-gate error (\(p_{\text{err}_H}\)),  $CX$-gate error (\(p_{\text{err}_{CX}}\)), and read-out error (\(p_{\text{err}_M}\)). For a node-to-Charlie distance of  \(D = 1 \, \mathrm{km}\), surpassing the performance of an unencoded Bell-state heralded protocol necessitates a significant reduction in gate error rates. Specifically, the gate error rates must be improved by an order of magnitude beyond what current experimental techniques can achieve  (\(p_{\text{err}_H}\)\(\lesssim 10^{-5}\), \(p_{\text{err}_{CX}}\)\(\lesssim 10^{-4} \), and \(p_{\text{err}_M}\)\(\lesssim 10^{-4}\)).  This result offers critical insight into the hardware improvements necessary to implement logical Bell state protocols with ion trap quantum memories. 
Under the physical noise model, code family thresholds were determined using gate error ratios defined as $\xi \times p_{e}$ with $p_{e}\in\{p_{\text{err}_H}, p_{\text{err}_{CX}}, p_{\text{err}_M}\}$, where \(\xi = 1.68 \pm 0.01\) for rotated Surface codes and \(\xi = 0.41 \pm 0.01\) for Bacon-Shor codes.

Additionally, Protocol 2 was analyzed in terms of its success rate showing rates up to \((32.53\pm1.53) \, \mathrm{Hz}\)  over distances of  $1$ to $80$ $\mathrm{km}$ between the end node and the intermediary node.

Having defined the hardware requirements for logical Bell state memory-assisted protocols, we explore future prospects and challenges associated with utilizing quantum error correction in quantum memories.
 



\section{Preliminaries}

\subsection{Notation}
This paper uses Dirac notation for quantum states and capital letters for quantum gates. General quantum error correction (QEC) codes are denoted as $ C[[n,k,d]] $, where  $ n $ is the number of data qubits, $ k $ is the number of logical qubits, $ d $ is the code distance and $C$ is the initial of the name of the code. Specific codes are represented as  $ S[[n,k,d]] $ for surface codes, $ BB[[n,k,d]] $  for bivariate bicycle codes, and  $ BS[[n,k,d]] $ for Bacon-Shor codes, providing a consistent framework for describing codes, subcodes, and low-density parity-check (LDPC) codes.


We introduce the unitary operators $X$ (Pauli-$X$) and $Z$ (Pauli-$Z$) by describing their action on the computational basis of $\mathcal{H}=\mathbb{C}^2$: $ X \ket{j} = \ket{j \oplus 1}$ and $Z \ket{j} = (-1)^{j}\ket{j}$. Similarly, we define the Hadamard gate $H$ and the CNOT gate $CX$ as:
\begin{equation}
    H \ket{j} = \frac{1}{\sqrt{2}} \Big( \ket{0}+ (-1)^{j}  \ket{1} \Big),
\end{equation}
\begin{equation}
    CX \ket{i,j}= \ket{i,i\oplus j},
\end{equation}
where $i$ and $j \in \{0, 1\}$. In summary, all quantum circuits adhere to the standard quantum computation formalism: time progresses from left to right and measurements for the QEC code are performed on a computational basis.

\begin{figure}
\includegraphics[width=0.5\textwidth,clip]{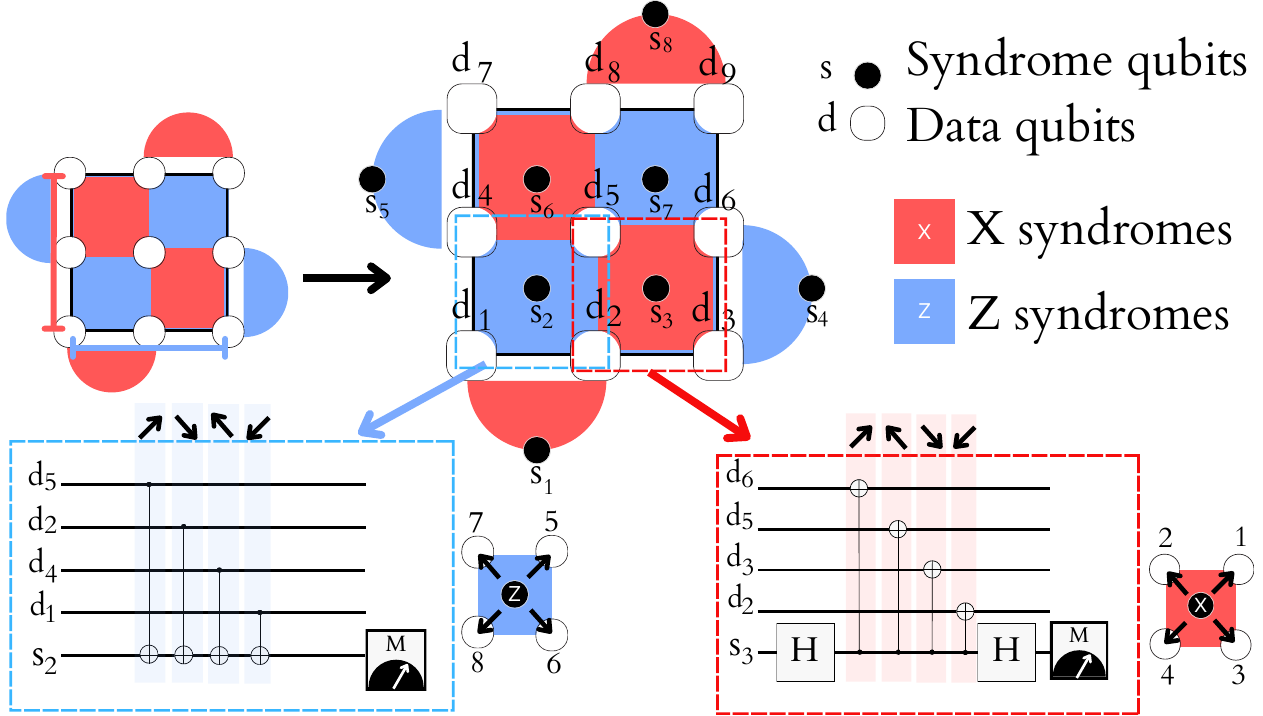}
\caption{Example of $d=3$ rotated surface code, with respective $Z$ and $X$ syndrome measurements. The set of X-stabilizers is associated with red plaquettes, while Z-stabilizers correspond to blue plaquettes. Since neighboring plaquettes always share two vertices, the stabilizers commute for any arrangement of plaquettes. The order of operations is presented with arrows between the X/Z syndrome qubits and data qubits.}
\label{surf}       
\end{figure} 
\subsection{Glossary}

For a general quantum error correction code $C[[n,k,d]]$ we introduce a new concept of code separability:

\begin{definition}
Code separability: 
A code $ C[[n,k,d]] $ is separable if it can be split into two independent codes: $C[[n,k,d]] \rightarrow (C_{1}[[n-s,k-a,d]], C_{2}[[s,a,d]])$, without compromising the original code's architecture or distance $d$. Where, code $C_{1}[[n-s,k-a,d]]$ consists of $n-s$ data and $k-a$ logical qubits, and $C_{2}[[s,a,d]]$ consists of $s$ data and $a$ logical qubits.
\end{definition}



An example of a separable code can be viewed in \cite{Horsman_2012} for the split operation using lattice surgery in surface codes (S): $S[[27,3,3]] \rightarrow (S_{1}[[18,2,3]],S_{2}[[9,1,3]])$. A generalization of such is presented in the context of Calderbank-Shor-Steane (CSS) code surgery \cite{Cowtan2024csscodesurgeryas}.

We also define the concepts of threshold and pseudo-thresholds error rate of a code family and code, according to \cite{svore2006}:

\begin{definition}
Threshold error rate: For a given noise model and code family, 
the threshold is the maximum physical error rate at which increasing the code distance (as \( d \to \infty \)) no longer results in an improvement in the logical error rate.
\end{definition}

\begin{definition}
Pseudo-threshold error rate:  
For a given noise model, the pseudo-threshold is the maximum physical error rate at which a specific code \( C[[n, k, d]] \) achieves the same logical error rate as the unencoded case.
\end{definition}

This work uses pseudo-threshold to compare the performance of unencoded Bell states with their logical counterparts.





\begin{figure}
\includegraphics[width=0.5\textwidth,clip]{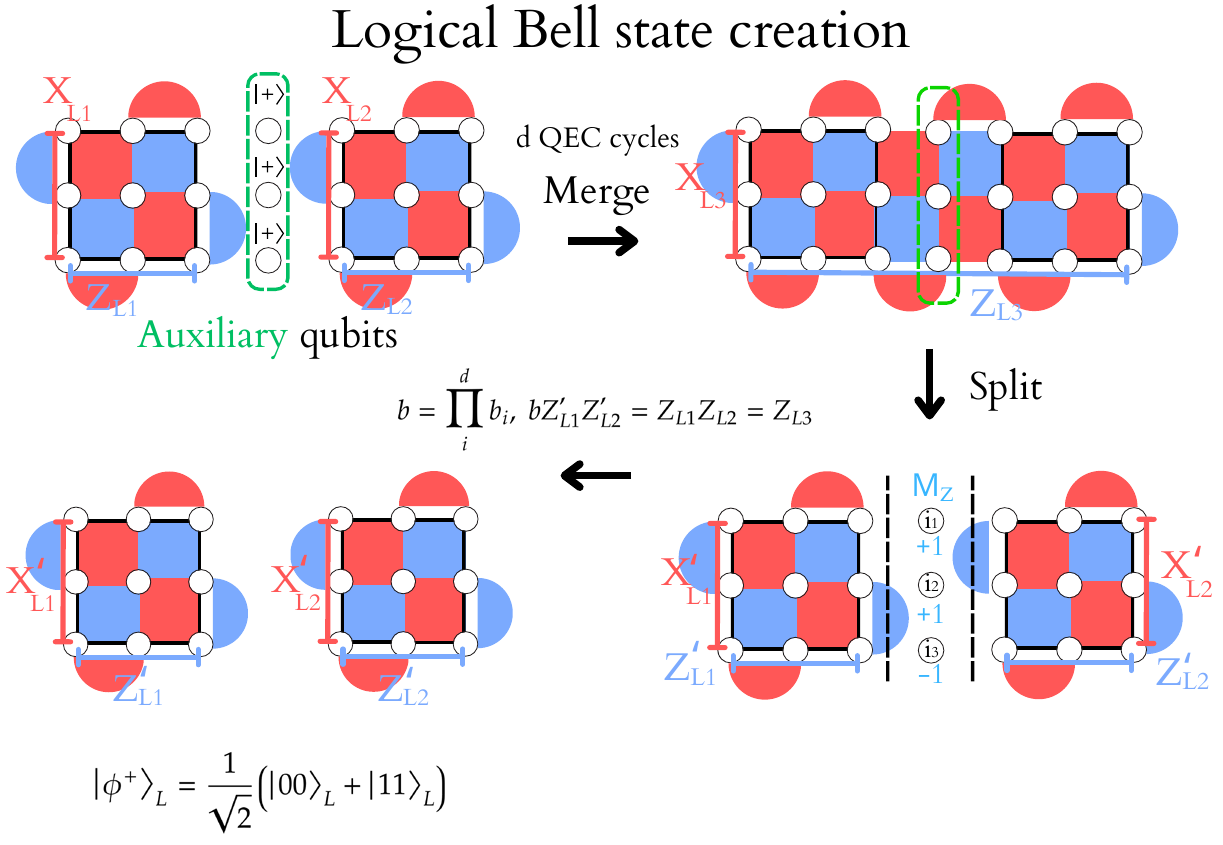}
\caption{Lattice surgery for creating logical Bell states by merging and splitting operations on a rotated surface code.}
\label{Bell}       
\end{figure}  
\subsection{Choice of QEC Codes}

We focus on the subsystem CSS Bacon-Shor code $ BS[[9,1,3]] $ and the CSS surface code $ S[[9,1,3]] $, both of which are well-suited for current hardware, enabling QEC with smaller circuit sizes and reducing noise errors. More specifically for $S$ codes, we consider both the planar lattice and a rotated one as described in \cite{Horsman_2012}. An example of rotated $S$ code for $d=3$ is demonstrated in Fig.\,\ref{surf}.
We also consider in this work $BS[[25,1,5]]$ and $S[[25,1,5]]$ codes, however, they are only used to calculate the code thresholds as the implementation of $BS[[25,1,5]]$ in the noise model described in Section\,\ref{robustn} is currently unfeasible due to the high qubit noise and loss. For a more indepth view of Surface codes and Bacon-shor codes operation see Appendix.\,\ref{Indepth_a}.

While bivariate bicycle codes ($BB$) are more scalable, their smallest configuration $BB[[18,4,4]]$ and $ BB[[18,2,2]]$ are not directly separable and hence not comparable with $ S[[9,1,3]] $ or $ BS[[9,1,3]] $ codes. Meaning we would need to generate a $ BB[[36,8,4]] $ or $BB[[36,4,2]]$ code to encode a logical Bell pair, the added logical qubits could be used to enhance redundancy or support logical Bell-pair distillation \cite{Xu2024, Cohen_2022}.  Additionally, the choice of quantum error correction codes must align with the physical constraints of the protocol. While toric codes \cite{Breuckmann_2021} and $BB$ codes \cite{Bravyi2024} offer potential benefits, their operations must be compensated in hardware due to repeated boundary conditions.

We also considered utilizing hybrid surface and Bacon-Shor codes, as discussed in \cite{bsc}. These codes are tailored to the IBM processor architecture and, for \( d = 3 \), yield the hybrid \( S|BS[[18,2,3]] \) code. Despite the code's strong adaptability to lattice architectures, its non-separability renders it incomparable to the \( d = 3 \) \( S \) and \( BS \) codes individually.

\subsection{Lattice surgery}

Lattice surgery is a technique that enables the splitting and merging of patches, thereby facilitating universal logical operations \cite{Horsman_2012, Litinski_2019}.  Each operation incurs a time cost proportional to the code distance $d$, with one unit of time approximately equating $d$  \cite{Litinski_2019}. Single logical qubit initializations can be achieved with states  $\ket{+}$, $\ket{-}$, $\ket{0}$ and $\ket{1}$ through transversal initialization, and states $\ket{i}$ and $\ket{-i}$ via topological initialization using twists. Arbitrary states can be approximated using Clifford + T gate circuits \cite{PhysRevLett.114.080502}, with no additional time cost for single qubit logical initializations.

Entanglement-based initialization requires the merging and splitting of patches, adding an extra time cost of $d$ cycles.
An example of logical Bell state generation is presented in Fig.\,\ref{Bell} for an $X$ boundary in a rotated $S[[18,2,3]]$ code.

A more in-depth explanation can found in \cite{Horsman_2012,Tan_2024,deBeaudrap2020zxcalculusis} but in summary $d$ auxiliary data qubits are initialized in the basis of the lattice surgery either $\ket{0}$ or $\ket{+}$, during $d$ QEC cycles the merging operation occurs. The merge operation realizes a $X_{L1}\otimes X_{L2}$ measurement, joining the boundaries between two codes.

Later on the split operation occurs, measuring the auxiliary qubits, the logical operations of the splitted codes $X^{'}_{L1}$ and $X^{'}_{L2}$ both commute the split operation. $Z_{L3}$ decomposes as $Z_{L3}=Z_{L1}Z_{L2}=bZ^{'}_{L1}Z^{'}_{L2}$ with an added correction $b$ as seen in Fig.\,\ref{Bell}, defined by the sign of the product of the measurements $b_{i}$ of the auxiliary qubits. After the codes are split and the correction is performed the logical Bell-pairs are generated $\ket{\phi^{+}}_{L}=\frac{1}{\sqrt{2}}(\ket{00}_{L}+\ket{11}_{L})$.
Lattice surgery for Bacon-Shor codes is performed in a similar manner, the strategy used is outlined in \cite{gidney2023baconthreshold}.

\begin{figure*}
\includegraphics[width=1\textwidth,clip]{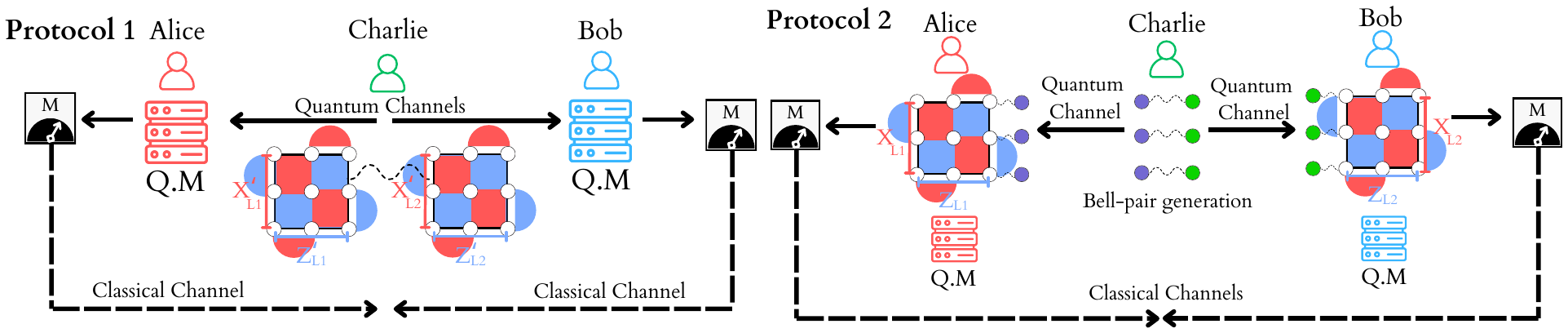}
\caption{Proposed protocols for QEC on a quantum memory for heralded entanglement protocol. Protocol 1 (left) and Protocol 2 (right) represent the local/non-local generation of logical Bell states, respectively. For Protocol 1 the data qubits are transmitted to Alice and Bob using a multi-mode channel and stored in a quantum memory (Q.M) while in Protocol 2 auxiliary Bell-pairs are generated entangling for $d$ cycles the boundaries of the codes already present inside the quantum memories.}
\label{protocol:LBPG}       
\end{figure*} 

\section{Logical Bell-pair protocols}
We propose to study two protocols as seen in Fig.\,\ref{protocol:LBPG}, a local (left) and a non-local (right) logical heralded entanglement, both of which leverage QEC and lattice surgery techniques in a multi-mode quantum storage to ensure robustness against quantum noise and operational errors. 

\subsection{Local logical Bell-pair protocol}
\begin{protocol}{Local logical Bell-pair protocol}
\begin{enumerate}
    \item Charlie constructs two QEC codes $C[[n,k,d]]$.   
    \item Using lattice surgery, Charlie merges each pair of codes and waits for $d$ QEC cycles.
    \item Charlie then splits the merged codes, and then transmits $n$ data qubits to Alice and Bob through a multi-modal quantum channel. 
    \item The protocol only continues if all data qubits are successfully gathered and stored in quantum memory; otherwise, it is aborted. 
    \item The logical qubits in the quantum memory undergo correction for $m_{1}$ cycles. Subsequently, the logical qubits $k$ are measured, and the results are communicated between both parties via a classical channel.
\end{enumerate}

\end{protocol}
The local protocol involves generating logical Bell-pairs locally by Charlie, taking advantage of the physical proximity of qubits to perform error correction and lattice surgery operations efficiently. Locally generated logical qubits are then shared with Alice and Bob over a quantum channel. This method typically involves splitting and merging logical qubits to create entangled states, with QEC codes ensuring that the logical qubits remain fault-tolerant throughout the process.

\subsection{Non-local logical Bell-pair protocol}

\begin{protocol}{Non-local logical Bell-pair protocol}
\begin{enumerate}
    \item Both Alice and Bob possess a $C[[n, k, d]]$ code in their quantum memory.
    \item Charlie generates $d$ auxiliary Bell-pairs during $d$ lattice surgery merging cycles and sends them through a single-mode quantum channel to Alice and Bob.
    \item If for one of the $d$ cycles, the auxiliary Bell-pairs are lost, the protocol continues and a newly generated auxiliary Bell-pair is sent.
    \item Alice and Bob expand the boundary of their respective codes using $d$ auxiliary Bell-pairs over $d$ cycles of syndrome measurement. They merge the two codes via entanglement using lattice surgery.
    \item Alice and Bob then split the merged qubits, resulting in the creation of two maximally entangled codes.
    \item The logical qubits in the quantum memory are corrected over $m_{1}$ cycles. Subsequently, the logical qubits $k$ are measured, and the results are shared between both parties via a classical channel.
\end{enumerate}
\end{protocol}

The non-local protocol focuses on generating logical Bell-pairs between spatially separated Alice and Bob. This approach extends the principles of QEC and lattice surgery to a distributed environment, where the boundaries of the codes are extended non-locally during $d$ merging cycles using auxiliary Bell states. Non-local logical Bell-pair generation requires advanced techniques to manage the additional challenges posed by noise and decoherence over long distances, but those challenges are heavily compensated by the high frequency of the auxiliary Bell state generation compared to the local scheme.


\subsection{Comparison of the two protocols}
Both protocols require \(d^2\) data qubits for Bell-pair production. Transitioning from Protocol 2 to Protocol 1 introduces certain hardware constraints. The non-local scheme (Protocol 2) operates within a single-mode optical channel, whereas the local scheme (Protocol 1) relies on multiplexing within an optical multi-mode channel for state transport.

The main benefit of the Protocol 1 is the early creation of fault-tolerant, maximally entangled logical pairs, reducing noise over $d$ cycles. Conversely, the Protocol 2 allows for resource-intensive entanglement purification at Alice’s and Bob’s nodes before encoding the Bell states, allowing even to perform purification for each of the $d^2$ auxiliary Bell-pairs depending on environmental noise. For instance, achieving a fidelity of \(F = 0.99\) from \(F = 0.85\) may require approximately \(5d^2\) auxiliary Bell-pairs, as extrapolated from the results in \cite{7010905}. This estimation assumes no loss of Bell-pairs during transportation or storage in the merging operation of Protocol 2.

In Protocol 1, the Bell-pair generation time is limited by the logical Bell-pair creation rate $ f_{\text{gen}} = 1/t_{\text{gen}}$ with $t_{\text{gen}}=t_{\text{merge}}+t_{\text{cycle}}$. If a photon or data qubits are lost due to conversion efficencies or non-demolition measurement efficencies, faulty detection dark counts, or fiber loss, the protocol is aborted and restarted. In Protocol 2, the auxiliary Bell-pair generation frequency matches the source frequency, $ f_{\text{source}}$. If any of the $d$ Bell-pairs fail during transportation or detection, we generate a new set of $d$ Bell-pairs for that merging cycle instead of restarting the entire protocol. 

In Protocol 2, the auxiliary Bell-pairs act as data qubits for merging the boundaries of the two codes, at Alice’s and Bob’s ends, requiring $d$ repetition cycles for each syndrome extraction step. Synchronization is vital between nodes, as Protocol 2 depends on continuous classical communication to exchange syndrome measurements. However, this constraint is partially relaxed in Protocol 1, where Charlie performs all lattice surgery operations prior to storing the states in quantum memory.



\section{Noise model}
\subsection{Depolarizing noise model}
\label{standard_noise_channel}
The depolarization channel \cite{PhysRevA.98.050301} is described by the following Kraus matrices for single-qubit and two-qubit channels 
\begin{eqnarray}
 K_{\mathrm{1}}=\{\sqrt{1-p_{\mathrm{err}}}I,\sqrt{\frac{p_{\mathrm{err}}}{3}}X,\sqrt{\frac{p_{\mathrm{err}}}{3}}Y,\sqrt{\frac{p_{\mathrm{err}}}{3}}Z\}, \\
 K_{\mathrm{2}}=\{\sqrt{1-p_{\mathrm{err}}}II,\sqrt{\frac{p_{\mathrm{err}}}{15}}XX, ... ,\sqrt{\frac{p_{\mathrm{err}}}{15}}ZZ\}.
\end{eqnarray}

In this model, the channel error probability is denoted as $p_{\mathrm{err}}$. The noisy channel is applied after (before) each gate (measurement) operation. Additionally we assume $p_{\mathrm{err}}$ for idle qubits during gate operation, differing from the model presented in \cite{PhysRevA.98.050301}. 

Additionally, this model does not account for decoherence and dephasing during the transportation and storage of quantum states.

For two-qubit parity measurements in the X and Z bases, denoted $ M_{XX} $ and $ M_{ZZ} $, the measurement process includes both the measurement operation and a two-gate depolarization channel. This depolarization channel, characterized by the Kraus matrix $ K_{\mathrm{2}} $, is applied to the qubits involved in the operation similarly to the method described in \cite{gidney2023baconthreshold}.

\begin{table*}
\caption{Simulation parameters used to describe the physical noise model.}
\centering
\begin{tabular}{|c|c||c|c||c|c| }
\hline
\textbf{Parameter} & \textbf{Value}  & \textbf{Parameter} & \textbf{Value} & \textbf{Parameter} & \textbf{Value}\\
\hline
Qubit decoherence time ($T_{1}$) & $3$ $ \mathrm{s}$ & Qubit dephasing time ($T_{2}$) & $0.5$ $ \mathrm{s}$ & Coupling strength ($g$) & $25\times2\pi$ $\mathrm{MHz}$  \\
\hline
Atomic dipole decay rate ($\gamma$) & $1.0\times2\pi$ $\mathrm{MHz}$ & Mode matching ($\mu_{\text{FC}}$) & $0.99 e^{-i0.03}$ & Total field decay rate ($\kappa$) & $27.8\times2\pi$ $\mathrm{MHz}$ \\
\hline
   Capture window ($t_{\text{rangeQ}}$) & $400$ $\mu\mathrm{s}$ & State transfer time & $100$ $\mu\mathrm{s}$ & QNDM time ($t_{\text{QNDM}}$) & $10$ $\mu\mathrm{s}$  \\
\hline
 QNDM eff. ($p_{\text{QNDM}}$)  & $0.855$ & State transfer eff. ($p_{\text{trs}}$)& $0.5$ & Fiber attenuation ($\tau$)& $0.17/0.7$ $\mathrm{dBkm}^{-1}$  \\
\hline
Source frequency ($f_{\mathrm{source}}$)  & $33$ $\mathrm{MHz}$
& Source wavelength & $1550$ $\mathrm{nm}$  &  Dark count probability ($p_{\mathrm{dark}}$)  & $0.03$   \\
\hline
 CX gate error ($p_{\mathrm{err_{CX}}})$ & $8.3\times10^{-3}$ &  H gate error ($p_{\mathrm{err_{H}}})$ & $2.1\times10^{-4}$
 & Readout error ($p_{\mathrm{err_{M}}})$  & $7.7\times10^{-3}$  \\

\hline
 Mean CX gate time & $970$  $\mu\mathrm{s}$  & Mean H gate time & $150$   $\mu\mathrm{s}$ & Mean measurement time  & $130$ $\mu\mathrm{s}$\\
\hline

\end{tabular}
\label{parameter:table} 
\end{table*}


\subsection{Physical noise model}
\label{robustn}

\subsubsection{Error model}
For the transmission channel, we assume for Protocol 1 a multi-mode fiber optic channel sized to the data qubits and for Protocol 2 a single-mode channel with source frequency of $f_{\mathrm{source}}=33$ MHz \cite{Liu2021}. The fiber loss channel is described by the single-photon transmission probability \cite{Coopmans_2021}
\begin{equation}
\eta_{\mathrm{channel}}=10^{-D\tau/10},
\label{fiber_eq}
\end{equation}
with $D$ being the transmission distance between the nodes (Alice and Bob) and Charlie, and $\tau=0.70 \,\mathrm{dBkm}^{-1}$ and $\tau=0.17 \,\mathrm{dBkm}^{-1}$ being the fiber attenuation for Protocol 1 (multi-mode) and 2 (single-mode) respectively, with a refractive index of glass in the fiber of $r_{\mathrm{glass}}=1.44$.

To track the photons before storage without destroying the encoded states in the data qubits we utilize quantum non-demolition measurements (QNDM) as modeled in \cite{Niemietz_2021}.

\begin{figure*}
\hspace{10mm}
\includegraphics[width=0.9\textwidth,clip]{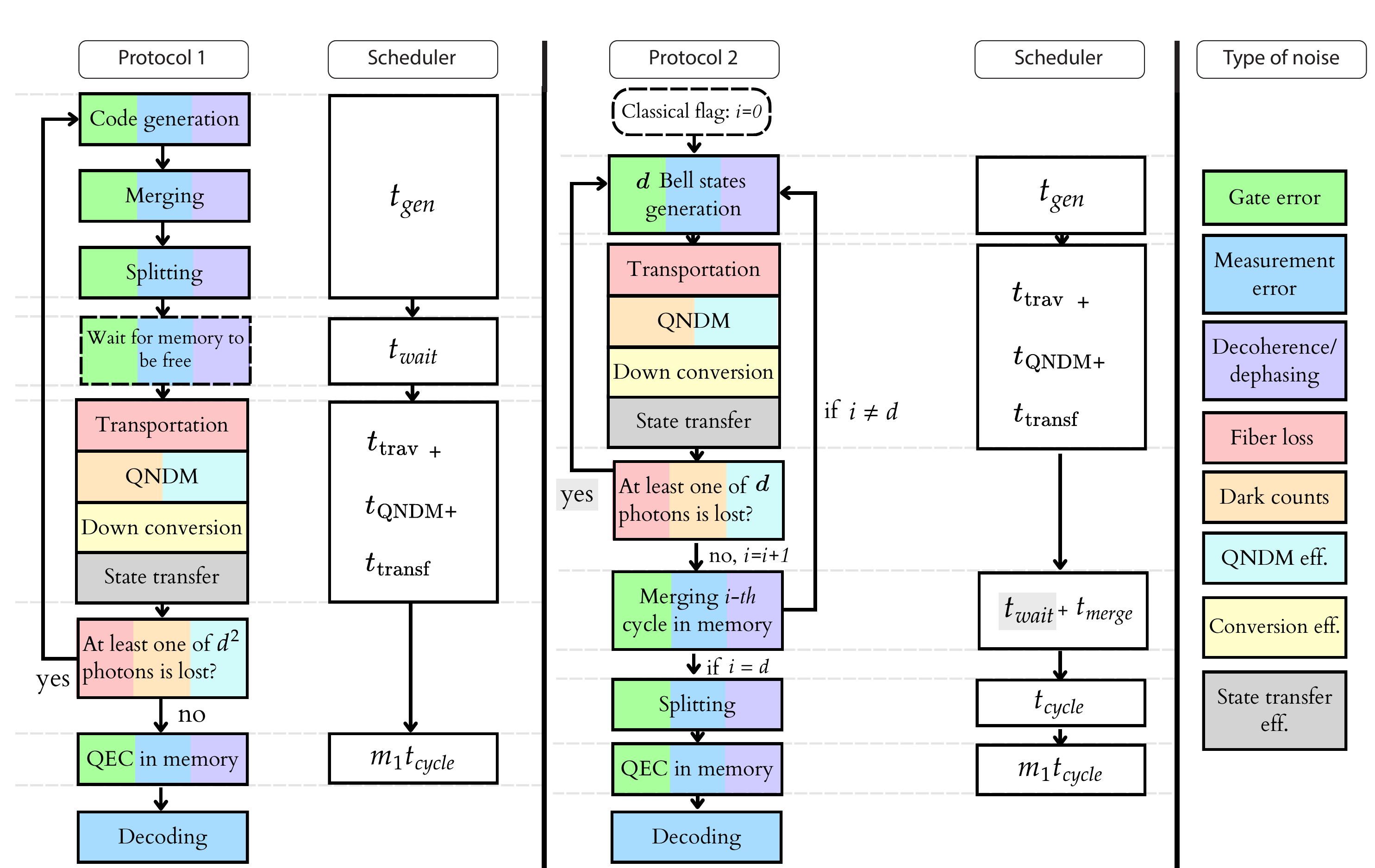}
\caption{Simulation overflow for both Protocol 1 and Protocol 2. The various noise levels, as detailed in Section\ref{robustn}, are represented using color gradients across each segment of the simulation. In parallel, the scheduler monitors the quantum memory occupancy and records the number of repetitions required whenever at least one of the $d$ photons is lost during the merging operation in Protocol 2.}
\label{simoverf}       
\end{figure*} 

The transmission channel model in Eq.\,\ref{fiber_eq} is modified to account for the detector efficiency. This modification considers the probability that an incoming photon $\ket{1_{iq}}$ successfully reaches the detector, as well as the subsequent efficiency of transmitting the photon $\ket{1_{oq}}$ given that a non-destructive detection event $\ket{0_{a}}$ has occurred:
\begin{equation}
\eta_{tot}=\eta_{\mathrm{channel}}P(0_{a}|1_{iq})P(1_{oq}|0_{a}).
\label{fiber_eq2}
\end{equation}

To perform quantum non-demolition measurements (QNDM), we consider a detector operating in a high cooperativity regime, defined by \(C = \frac{g^2}{2\gamma \kappa} > 1\), where \(\gamma\) represents the atomic dipole decay rate, \(g\) is the coupling strength with the qubit cavity mode, and \(\kappa\) is the cavity decay rate. Since none of the observed losses appear to be fundamental \cite{Reiserer_2013}, we theoretically model the optimal experimental parameters to adapt them to our protocols using input-output theory \cite{Kuhn2015}.  

We simulate the system using QuTiP \cite{Johansson_2012} in accordance with \cite{Niemietz_2021} with a more detailed explanation in Appendix\,\ref{Ap_dec}. Furthermore, we ensure that the QNDM readout time aligns with the emission frequency of the data qubit in Protocols 1 and 2, considering an added readout time of \(t_{\text{readout}} = 1 \, \mu s\) for a typical \(\pi/2\) atomic manipulation pulse \cite{Hacker2016}. By improving the atomic dipole decay rate to \(\gamma = 1 \times 2\pi\) MHz, adjusting the cavity field decay rate contribution into the reflection mode to \(\kappa_{\text{r}} = \frac{4.0}{4.3}\kappa\), and achieving a near-perfect fiber-cavity mode matching amplitude of \(\mu_{\text{FC}} = 0.99 e^{-i0.03}\) compared to the results in \cite{Niemietz_2021}, we obtain a transmissivity of \(P(1_{oq}|0_a) \approx 0.90\). For \(P(0_{a}|1_{iq})\), current QNDM efficiencies range from \(0.45 \pm 0.02\) \cite{Niemietz_2021} to \(0.74\) \cite{Reiserer_2013}. With the parameters considered, we can achieve \(P(0_{a}|1_{iq}) \approx 0.95\). We assume that any loss in fidelity after non-destructive detection of photonic qubits is negligible. The dark count contribution to the overall signal transmission is approximately \(p_{\text{dark}} = 0.03\).

\subsubsection{Frequency conversion} 
As the frequency from ion trap memories belongs in the visible range \(\lambda_a = 750 - 855 \text{nm}\) a down conversion from telecom frequency is needed.
We analyzed the vital parameters to perform the difference frequency generation (DFG) in ridge waveguide.  
Theoretically the frequency conversion can be described by coupled mode equations governing three-wave mixing
in waveguides \cite{Zaske_2013}. With coupling constants ($\kappa_1/\omega_{1}=\kappa_2/\omega_{2}=\kappa^*_3/\omega_{3}\equiv\kappa_{f}$) depending on the spatial three-mode overlap they have a general form of \cite{Zaske_2013} 

\begin{equation}
\kappa_1 = \frac{\omega_1 \varepsilon_0 d_Q}{2} \int \int E_{1x}^*(x, y) E_{3x}(x, y) E_{2x}^*(x, y) \, dx dy.
\label{loss}
\end{equation}

Here, \(\omega_1\) is the frequency of the field, \(E_m(x,y)\) represents the normalized transverse intensity distribution, \(\varepsilon_0\) is the vacuum permittivity, and \(d_Q\) is the nonlinear coefficient. 

The conversion efficiency is defined as \cite{Roussev:04}:

\begin{equation}
\eta_{conv} = \eta_{max} \sin^2\left(\sqrt{|\kappa_{f}|^{2} \omega_a \omega_b P_p} L\right).
\label{loss2}
\end{equation}

In this expression, \(\eta_{max}\) accounts for potential attenuation losses, while \(\omega_a\) and \(\omega_b\) correspond to the input (\(\lambda_a = 750 - 855 \text{nm}\)) and output (\(\lambda_b = 1550 \text{nm}\)) frequencies, respectively. Additionally, \(P_p\) denotes the pump power, and \(L\) is the waveguide length. 

In the absence of losses (\(\eta_{max} = 1\)), complete conversion is theoretically possible without intrinsic limitations. For \(P_p = 100 \text{mW}\) and \(L = 40 \text{mm}\), the conversion efficiency exceeds \(0.95\).  

Experimentally, conversion efficiencies from telecom wavelengths to ion-trap frequencies have been demonstrated to range between \(0.265\) \cite{Bock_2018} and \(0.35\) \cite{saha2023lownoisequantumfrequency, Krutyanskiy2017}. One major source of loss is the presence of higher-order spatial and axial modes beyond the three-mode interaction described in Eq.\,\ref{loss}, which remain largely unconverted when populated.

In \cite{Bock_2018}, it was argued that increasing the Bragg grating efficiency and implementing optimized dichroic mirrors could push the conversion efficiency beyond \(0.5\). In \cite{Krutyanskiy2017}, if in-coupling aspheres, out-coupling aspheres, shortpass filters, dichroic mirrors, and longpass filters are excluded from consideration, a conversion efficiency of \(0.59\) is obtained. Moreover, assuming equal waveguide transmission losses at \(1550 \text{nm}\) and \(750-855 \text{nm}\), the conversion efficiency is projected to reach approximately \(\eta_{conv}\sim0.9\). We adopt this value for our analysis and consider any added delay times from the conversion negligible. 

\subsubsection{Quantum memory}

At both Alice's and Bob's locations, we utilize ion-trap quantum memories, which have been experimentally demonstrated as quantum network nodes \cite{PhysRevLett.130.090803}. These memories have been used in quantum networks \cite{PhysRevLett.130.213601}, with ion-photon entanglement observed over distances of up to 101 km \cite{PRXQuantum.5.020308}.

For quantum error correction, we require quantum memories that support state manipulation while the quantum state is stored. Additionally, coherent quantum transfer of the photon state to the ion memory must be achieved while preserving any prior system entanglement. The gate and measurement errors were derived from the IonQ Forte processor calibration data. For decoherence and dephasing, we consider values on the order of seconds, with \( T_1 = 3 \text{s} \) and \( T_2 = 0.5 \text{s} \) .

\begin{table*}[h!]
\caption{Time budget for one QEC cycle at the quantum memory for $D=1$  $\mathrm{km}$.}
\centering
\begin{tabular}{|l c c c c|} 
\hline
 QEC Code & $t_{\text{cycle}} (\mathrm{s})$ & $t_{\text{merge}}(\mathrm{s})$  & $t_{\text{trav}}(\mathrm{s})$ & $t_{\text{total}}(\mathrm{s})$ \\  
 \hline 
 Protocol 1: $S[[18,2,3]]$ (incl. rotated) & $4.31\times10^{-3}$ & $1.72\times10^{-2}$ & $4.80\times10^{-6}$ & $2.60\times10^{-2}$   \\
 Protocol 1: $BS[[18,2,3]]$ & $8.88\times10^{-3}$ & $3.55\times10^{-2}$ & $4.80\times10^{-6}$ & $5.34\times10^{-2}$  \\
 Protocol 2: $S[[18,2,3]]$ (incl. rotated) & $4.31\times10^{-3}$ & $1.72\times10^{-2}$ & $1.44\times10^{-5}$ & $2.60\times10^{-2}$ \\
 Protocol 2: $BS[[18,2,3]]$ & $8.88\times10^{-3}$ & $3.55\times10^{-2}$ & $1.44\times10^{-5}$ & $5.34\times10^{-2}$   \\
 \hline
\end{tabular}
\label{time_budget}
\end{table*}

More recent studies \cite{PRXQuantum.2.020331} have integrated ion traps with optical cavities, significantly enhancing photon capture efficiency to $0.72$, with a theoretical maximum of $0.84$. This improvement has led to the generation and detection of ion-entangled, fiber-coupled photons with an efficiency of $0.426$ \cite{PRXQuantum.2.020331}. Given these advancements, and assuming negligible optical losses, we adopt an optimistic photon-to-ion state transfer efficiency of \(\eta_{trs} = 0.5\), we consider an added time for capture and state transfer of the photons of $100$ $\mu\mathrm{ s}$. 


\section{Results}
Using Stim \cite{Gidney2021stimfaststabilizer}, we simulated the performance of both protocols for planar and rotated surface codes with $d=3$ and $d=5$, as well as Bacon-Shor codes, under both the depolarization and physical noise models. The simulations were conducted using the Monte Carlo method with $10^{8}$ events for the depolarization channel and $10^{5}$ events for the physical noise model, the simulation overflow for Protocols 1 and 2 can be found in Fig.\,\ref{simoverf}. To decode $X$, $Y$, and $Z$ errors, we employed a minimum-weight perfect matching algorithm implemented via the sparse blossom method, as detailed in \cite{higgott2023sparseblossomcorrectingmillion}. 

\subsection{Depolarizing channel}

\subsubsection{Logical error rate}

\begin{figure}
\includegraphics[width=0.45\textwidth,clip]{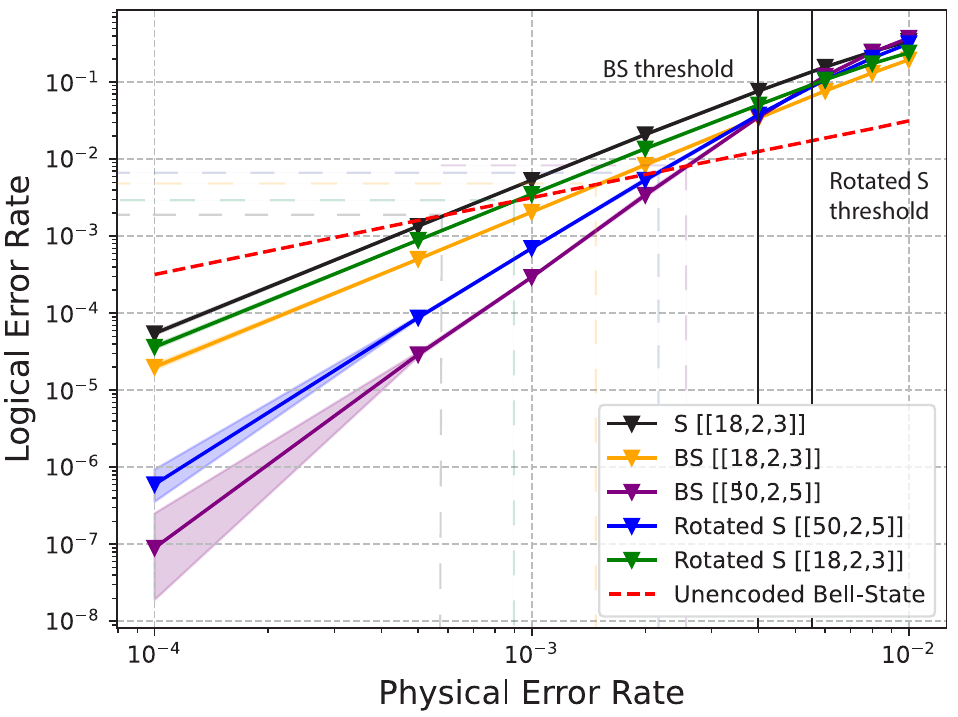}
\caption{An analysis of one iteration of quantum error correction (QEC) for logical Bell-pairs stored in quantum memory is conducted using the standard noise model described in Section\,\ref{standard_noise_channel}. The study determines the thresholds where increasing the code distance (from $d=3$ to $d=5$) results in a higher logical error rate for both the $BS$ code and the rotated $S$ code. Additionally, pseudo-thresholds are identified, representing the noise levels at which QEC becomes advantageous. This is established by comparing the logical error rates of the codes to those of an unencoded Bell state.}
\label{1QEC}       
\end{figure}

Under the depolarization channel, we examined in Fig.\,\ref{1QEC} how physical error rates after each operation affect logical error rates over a single QEC iteration for logical Bell-pairs in quantum memory. We observe pseudo-thresholds where physical error rates above  $p_{\text{err}} = (5.5\pm 0.2)  \times 10^{-4}$ for $S[[18,2,3]]$, $p_{\text{err}} = (9.0\pm 0.3)  \times 10^{-4}$ for rotated $S[[18,2,3]]$, and $p_{\text{err}} = (1.5\pm 0.2) \times 10^{-3}$ for the $BS[[18,2,3]]$ codes show no advantage over unencoded Bell states. This extends the results of \cite{PhysRevA.98.050301}, where pseudo-thresholds of $p_{\text{err}} = 9.0\times10^{-3}$ for Bacon-Shor and $p_{\text{err}} = 1.5\times10^{-3}$ for surface codes were obtained without lattice surgery. Notice that compared to \cite{PhysRevA.98.050301} our standard model considers depolarization for idle gates. These results apply to both protocols, as the standard noise model does not introduce significant differences between them.

\begin{figure}
\includegraphics[width=0.45\textwidth,clip]{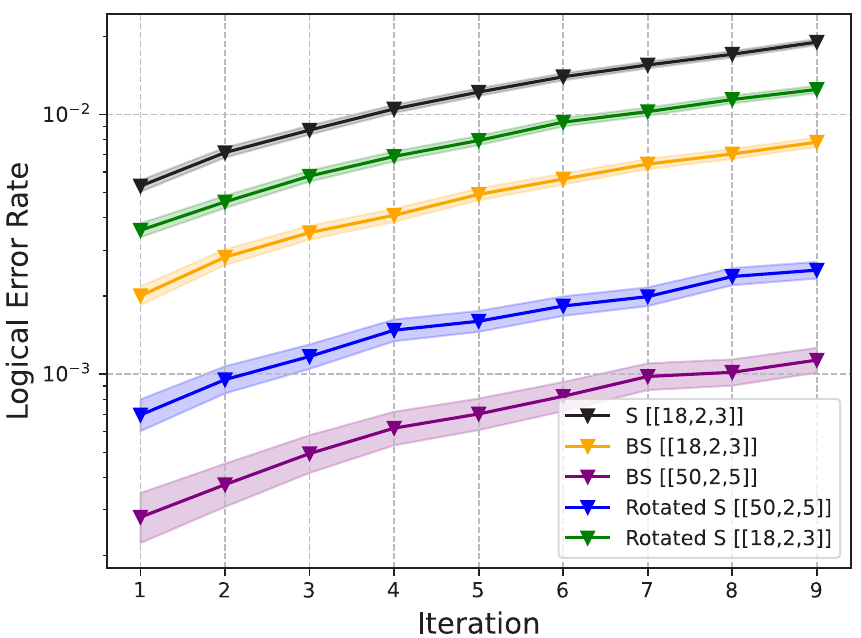}
\caption{The analysis involves 10 iterations of quantum error correction (QEC) applied to logical Bell-pairs stored in quantum memory, utilizing the depolarizing noise model with a parameter value of $p_{\text{err}}=10^{-3}$.}
\label{20QEC}       
\end{figure}

In Fig.\,\ref{20QEC}, we additionally studied the overall performance of the codes for $p_{\text{err}}=10^{-3}$, where $BS$ code outperformed both surface codes for 9 QEC syndrome extractions in the quantum memory, reaching for $d=3$ a logical error of $p_{L}=(7.8\pm0.3)\times10^{-3}$. Despite the standard noise model's usefulness for evaluating per gate operation errors, it does not account for photon loss during transportation and storage, and incorrect photon detection, requiring a more robust evaluation of code viability.

\subsubsection{Code scalability}

To assess the scalability of the protocol in relation to code performance, we identified the threshold under a specific error model in which increasing the code distance results in a deterioration of the logical error rate. This was achieved by comparing the logical error rates of codes with distances \(d=3\) and \(d=5\) as a function of \(p_{\text{err}}\). 

For the standard noise model described in Section\,\ref{standard_noise_channel}, the thresholds were identified in Fig.\,\ref{1QEC} as $p_{\text{err}} = $\((3.9 \pm 0.1) \times 10^{-3}\) for \(BS\) codes and $p_{\text{err}} = $\((5.8 \pm 0.2) \times 10^{-3}\) for rotated \(S\) codes.


\subsection{Physical noise model}
\begin{figure*}[t]
\includegraphics[width=1\textwidth,clip]{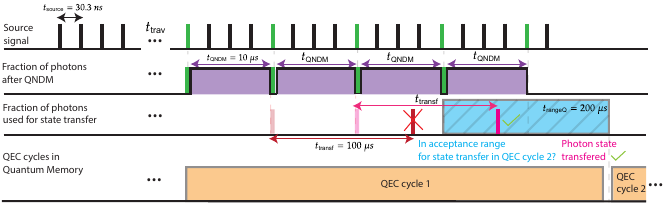}
\caption{
The scheduler monitors the number of auxiliary Bell-pairs that are ready for the next QEC iteration. Due to the limited capture efficiency of QNDM, only a subset of the available Bell-pairs is registered. Specifically, if a photon survives the state transfer processing time \(t_{transf}\) and remains within the acceptable time window \(t_{rangeQ}\), its state is successfully transferred into quantum memory for use in the next QEC cycle.}
\label{img:Scheduler}       
\end{figure*}

\begin{figure*}
\includegraphics[width=1\textwidth,clip]{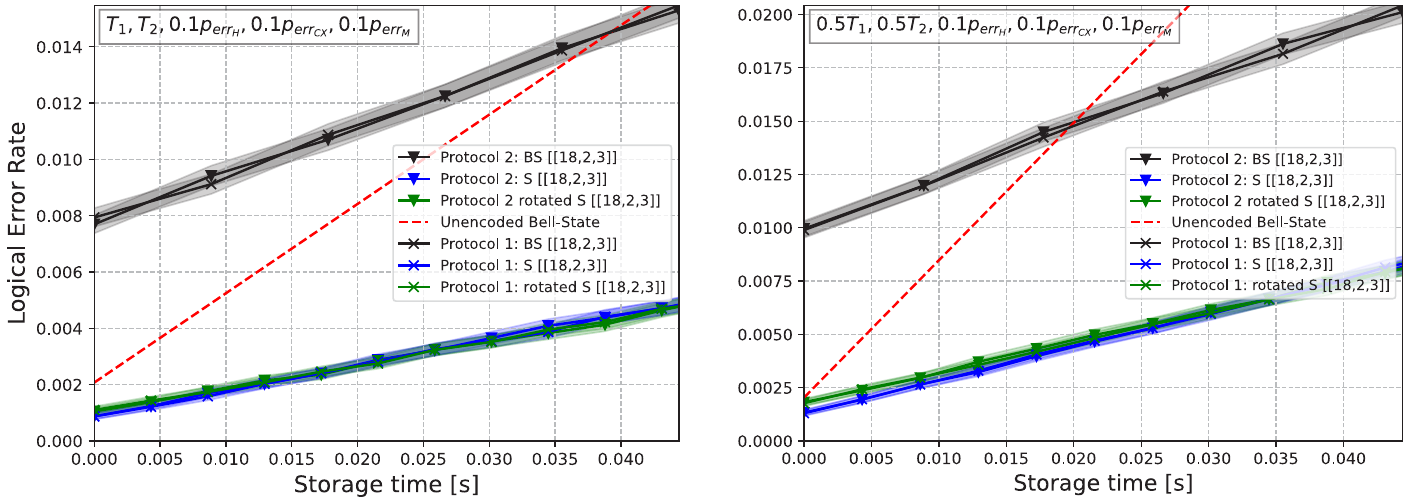}
\caption{The time dependency for Surface and Bacon-Shor codes with a code distance of $d=3$ is analyzed, considering the effects of decoherence times $T_{1}$ and dephasing times $T_{2}$ with overall gate error $p_{\text{err}_{H}}, p_{\text{err}_{CX}}, p_{\text{err}_{M}}$, as outlined in Table \ref{parameter:table}. To determine a feasible threshold for implementing quantum error correction (QEC) for each code, these parameters were systematically varied.
Each point on the plot corresponds to an additional QEC cycle within the storage time framework for the respective code, providing a clear visualization of how the error rate evolves with successive iterations over time.}
\label{standard_noise3}       
\end{figure*}

\subsubsection{Logical error rate}
Using the error model from Table \ref{parameter:table}, we calculate the time budget in Table \ref{time_budget} for each protocol and various QEC codes to perform one QEC iteration cycle (\(t_{\text{QEC}}=t_{\text{cycle}}\)) for the logical Bell states in the quantum memory at $D=1$ $\mathrm{km}$.

For both protocols, we developed a scheduler to monitor the availability of quantum memories and the hardware responsible for logical Bell state generation. Fig.\,\ref{img:Scheduler} depicts the scheduler logic for Protocol 2, which manages the acceptance of auxiliary Bell-pairs before each QEC cycle during the merging process. Prior to each merging cycle, we define an acceptance window of \( t_{rangeQ} = 400\mu s \) to acquire the required \( d \) auxiliary Bell-pairs and facilitate state transfer. This selected acceptance window enables the collection of an average number of successful Bell-pairs ranging from \( 5.28 \pm 3.78 \) at $ 1 $ $\mathrm{km}$ to \( 2.26 \pm 2.41 \) at  $10$ $\mathrm{km}$ per merging cycle. These statistics were derived from 50 distinct merging cycles. 
During QNDM, however, only a single photon is accepted due to the lack of simultaneous multi-photon QNDM capability. In contrast, we assume that simultaneous state transfer into ion traps is possible, which would only contribute an overall delay in the state timing.

We adopt an optimal approach, keeping Charlie and the nodes synchronized through continuous classical communication. In Fig.\,\ref{standard_noise3}, for $10^{5}$ events per iteration, we simulated all code and protocol combinations, including the unencoded Bell state, analyzing simulation parameters $T_{1}$ and $T_{2}$ and the overall error rate $p_{\text{err}}$ for a distance of $D = 1$ $\mathrm{km}$.


Instead of comparing codes based on syndrome measurements, we evaluate them in terms of storage time. This approach simplifies comparison due to differences in operational complexity among codes, as discussed in Appendix.\ref{multimeasure}. Because of the time requirements for  $H$, $CX$, and measurement operations, a single QEC cycle of the Bacon-Shor code allows up to two syndrome extractions within surface codes in the same period of time.

In Fig.\,\ref{standard_noise3}, we estimate the pseudo-threshold at which QEC becomes more advantageous than using the unencoded Bell state. 
For $H$-gate error \(\lesssim \mathcal{O}(10^{-5})\), $CX$-gate error \(\lesssim \mathcal{O}(10^{-4}) \), and read-out error \(\lesssim \mathcal{O}(10^{-4})\), both rotated and planar surface codes offer lower error rates compared to the unencoded case. 
A similar trend is observed for Bacon-Shor codes with multiple iteration cycles, although they face challenges due to the complexity of initializing logical Bell states. This result is crucial as it defines the minimal hardware requirements under the stated assumptions for which QEC is viable in ion trap memories for logical Bell state protocols.


\begin{figure}
\includegraphics[width=0.45\textwidth,clip]{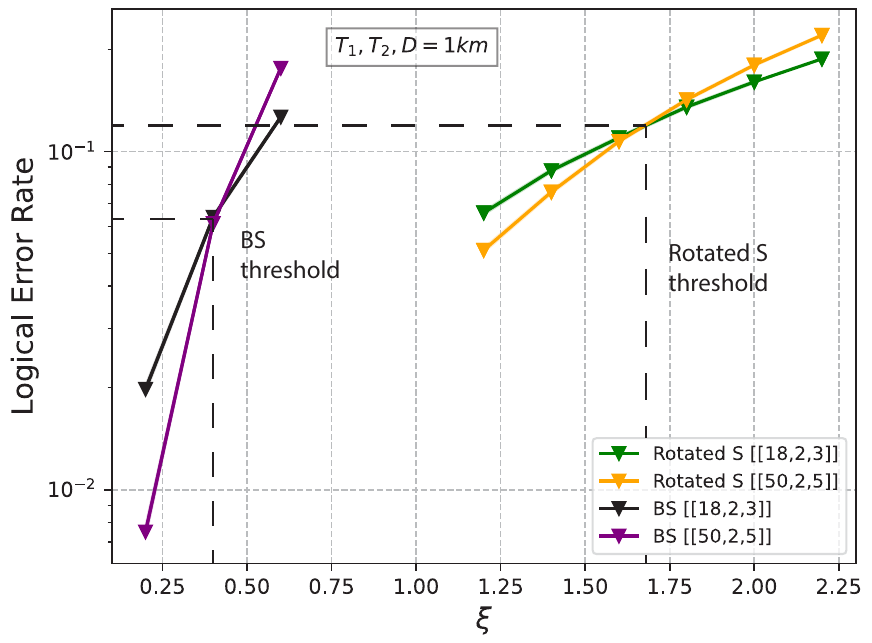}
\caption{We examine a single iteration of quantum error correction (QEC) applied to logical Bell-pairs stored in quantum memory, using the physical noise model detailed in Section\,\ref{robustn}. Thresholds for the code families, specifically the rotated S and BS codes, are identified based on the gate error ratio $\xi \times p_{e}$ (for $p_{e} \in \{p_{\mathrm{err_H}}, p_{\mathrm{err_{CX}}}, p_{\mathrm{err_M}}\}$, as defined in Table \ref{parameter:table}), assuming a distance of \( D = 1 \) km and using \( T_1 \) and \( T_2 \) from Table \ref{parameter:table}.}
\label{real_thresh}       
\end{figure}

\begin{figure}
\includegraphics[width=0.45\textwidth,clip]{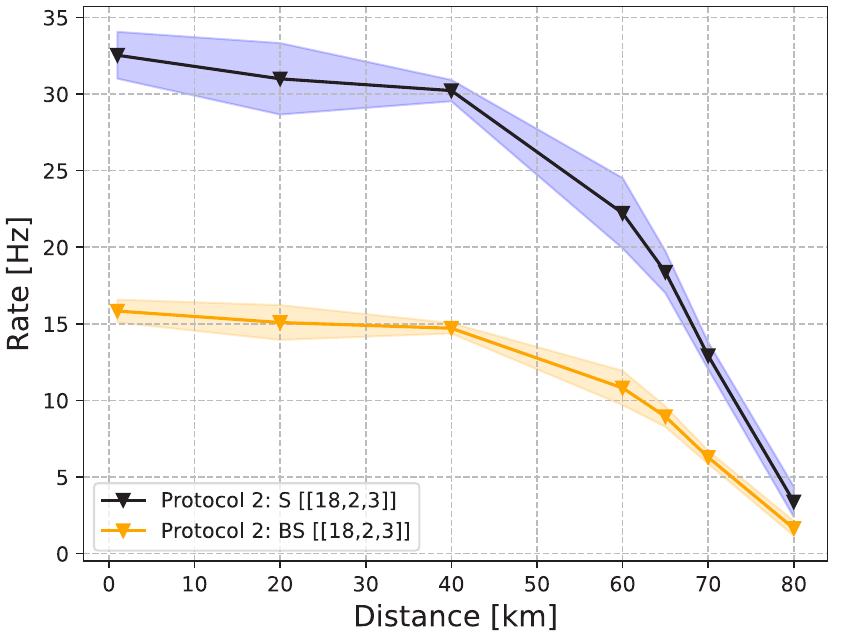}
\caption{Protocol rate in terms of operational distance between Charlie and the nodes for one QEC cycle of the logical Bell-pair in the quantum memory.}
\label{rate}       
\end{figure} 

\subsubsection{Code scalability}

For the noise model defined in Section \ref{robustn}, we determine the code family threshold in Fig.\,\ref{real_thresh}. This is done by evaluating the ratio of gate errors $\xi \times p_{e}$ with $p_{e} \in \{p_{\mathrm{err_H}} ,  p_{\mathrm{err_{CX}}} , p_{\mathrm{err_M}}\}$ at which increasing the code distance no longer reduces the logical error rate for a single QEC cycle, assuming \( D = 1 \) km and the \( T_1 \) and \( T_2 \) parameters from Table \ref{parameter:table}. For rotated \( S \), we find the threshold at a ratio of \( \xi=1.68 \pm 0.01 \). For \( BS \) codes, the high error per QEC cycle, discussed in Appendix \ref{multimeasure}, leads to a lower threshold of \(\xi= 0.41 \pm 0.01 \).

\subsubsection{Protocol rate}

In Fig.\,\ref{rate}, we illustrate the success rate of a single QEC cycle while the logical Bell states are stored in the memories of Alice and Bob. Protocol 1 is omitted from this analysis due to its inherent limitations, including prolonged gate processing times and the necessity to abort operations when a data qubit is lost, making it less practical.  On the other hand, Protocol 2 demonstrates performance, achieving success rates of up to \((32.53 \pm 1.53) \, \mathrm{Hz}\) over distances ranging from $1$ to $80$ $\mathrm{km}$ between the end node and the intermediary node.

The study emphasizes the performance of the \( S[[18,2,3]] \) and \( BS[[18,2,3]] \) error-correcting codes. These codes were analyzed over a sequence of $6.0 \times 10^{5}$ auxiliary Bell-pair generations. The standard deviation is determined based on five repetitions using different sampling seeds. To facilitate efficient state transfer during each \( d \) merging cycle, an acceptance window of \( t_{\text{rangeQ}} = 400 \, \mu\mathrm{s} \) was employed. This window optimized the interaction between the \( d \) auxiliary Bell-pairs and the corresponding ions, ensuring the success of the process.

\section{Conclusions}
This study aims to bridge the existing research gap by conducting detailed simulations to generate logical Bell-pairs using Quantum Error Correction (QEC) techniques. QEC is critical in second and third-generation quantum repeaters for maintaining state integrity during distillation and entanglement swapping. For the first time, we explore the feasibility of performing and storing logical Bell states through two logical heralded entanglement protocols: one local and one non-local.  Our comprehensive analysis delves into the physical parameters, minimal requirements, and challenges associated with generating and storing logical Bell-pairs within ion trap memories. We assess rotated and planar surface codes alongside Bacon-Shor codes, which can be implemented with low code distances considering existing hardware.

Two noise models are considered for logical heralded entanglement: a standard noise model (Section \ref{standard_noise_channel}) and a more realistic physical noise model (Section \ref{robustn}). We identified the pseudo-thresholds where exceeding the physical error offers no advantage over unencoded Bell states in either protocol. By comparing $d=3$ and $d=5$ codes we have also established thresholds beyond which increasing code distances degrade logical error rates (\(p_L\)). To evaluate the feasibility of the protocol, we studied the non-local protocol, which is the more practical of the two, achieving rates of up to \((32.53\pm1.53)\,\mathrm{Hz}\) over distances ranging from \(1\) to \(80\) km between the end node and the intermediary node.

We reevaluated the pseudo-thresholds in terms of gate error rates \(p_{\text{err}_H}\), \(p_{\text{err}_{CX}}\), and \(p_{\text{err}_M}\), with parameters provided in Table\,\ref{parameter:table}. For a node-to-Charlie distance of \(D = 1 \, \mathrm{km}\), achieving an advantage over an unencoded Bell state heralded protocol requires reducing gate error rates by an order of magnitude (\(p_{\text{err}_H}\)\(\lesssim 10^{-5}\), \(p_{\text{err}_{CX}}\)\(\lesssim 10^{-4} \), and \(p_{\text{err}_M}\)\(\lesssim 10^{-4}\)) 
These results underscore the significant hardware improvements needed to implement logical Bell state protocols in quantum memories. 

Future research could investigate leveraging unused logical qubits in quantum error-correcting codes to enhance redundancy and improve the fidelity of logical states. For higher number of data qubits, comparing $BB$ codes to $S$ and $B$ codes might further highlight their scalability advantages and challenges of implementing them in a quantum network scenario. However, this approach first requires addressing the challenges posed by cyclic boundary conditions and accounting for the memory's coupling map. These challenges could be mitigated through equivalent and supplementary operations implemented directly in the hardware \cite{Beverland_2022}. Addressing cyclic boundary conditions, potentially through hardware modifications like using Rydberg atoms \cite{pldpc}, remains an important step toward the practical implementation of these codes in quantum repeater protocols.


\section{Code availability}
All codes responsible for the results in this article can be found at: \href{https://github.com/terrordayvg/Logical-Bell-States}{https://github.com/terrordayvg/Logical-Bell-States}.

\section{Acknowledgments}
VG and JN acknowledge the financial support by the Federal Ministry of Education and Research of Germany in the programme of “Souverän. Digital. Vernetzt.”. Joint project 6G-life, project identification number: 16KISK002 and via projects 16KISQ039, 16KISQ077, 16KIS1598K, 16KISQ093 and  by the DFG via project NO 1129/2-1. 
VG also extends gratitude to Pol Julià Farré for reviewing the earlier versions of the article.

\bibliographystyle{IEEEtran}
\bibliography{bibliography_up}

\section{Appendix}

\subsection{Indepth view of QEC codes}
\label{Indepth_a}
\subsubsection{Surface codes}

Surface codes encode $d^2$  data qubits into a single logical qubit, requiring  $d^2 - 1$ syndrome qubits to measure  $X$ and $Z$  syndromes per error correction cycle (Fig. \ref{surf}), where $d$ is the code distance which corresponds to the maximum length of error chains that can be reliably detected and corrected by the decoding algorithm. 

Logical operations $X_L$ and $Z_L$ are implemented as products of Pauli operators, connecting opposite boundaries of the code. During each quantum error correction cycle, syndromes are measured for adjacent data qubits in a specific sequence with arrows, alternating between $X$and $Z$ stabilizer measurements as (Fig. \ref{surf}). When an error occurs, the corresponding measurement outcome flips its sign. If fewer than $(d-1)/2$ errors occur, each measured syndrome can be associated with a correction operation. A decoder \cite{deMartiiOlius2024decodingalgorithms} then estimates the error and applies a correction, restoring the system to the stabilizer $+1$ state without introducing a logical error.

Surface codes offer a high threshold of $1\%$, rapid decoding algorithms, and compatibility with two-dimensional lattice architectures in quantum processors. Experimental studies, such as Acharya et al. (2023) \cite{Acharya2023}, demonstrate their effectiveness, showing better logical error performance for distance-5 surface codes compared to averaged distance-3 surface codes in terms of logical error probability and error per cycle.

\subsubsection{Bacon-Shor codes}
Bacon-Shor codes \cite{PhysRevLett.98.220502} work similarly as surface codes, however they are built on a $L\times L$ planar lattice where $2(L-1)$ weight-$2L$ stabilizers are inferred from measurements of $(L-1)^{2}$ weight-2 gauge operators built as two-qubit parity measurements as seen in \cite{PhysRevA.98.050301}.  Each horizontal and vertical connection between qubits corresponds to $X\otimes X$ and $Z\otimes Z$ measurements, respectively. At each quantum error correction cycle the total parity of all data qubits is measured in Z and X basis. 

Bacon-Shor codes enhance protection against noisy channels by encoding logical qubits with non-commuting gauge operators. They offer benefits in qubit requirements and processing times for small distances but lack scalability to very low error rates \cite{gidney2023baconthreshold}. Experimental implementations are detailed in \cite{egan2021d} and \cite{PhysRevA.98.050301}.


Most experimental QEC efforts focus on logical single qubits due to current hardware limitations, including noise levels and decoding complexities. Techniques for logical multi-qubit operations include transversality \cite{PhysRevA.99.022330}, braiding \cite{Raussendorf_2007, Fowler_2012}, gate teleportation \cite{Gottesman1999}, magic state distillation \cite{PhysRevA.71.022316}, and lattice surgery \cite{Horsman_2012,deBeaudrap2020zxcalculusis}, with lattice surgery being preferred for its lower resource overhead.

\subsection{Two-qubit X and Z basis parity measurement}
\label{multimeasure}
To define the noise model for the two-qubit $ X $ or $ Z $-basis parity measurement ($ M_{ZZ} $ and $ M_{XX} $ respectively) operations in the context of the ion trap, we consider an equivalent decomposition into $ H $, $ CX $, and $ Z $-basis measurement operations as no native $M_{ZZ}$ and $M_{XX}$ gates are available for the IonQ processor. The decomposition for $ M_{XX} $, which measures parity in the $ X $-basis, can be expressed as follows (up to a global phase):

\begin{equation}
\begin{quantikz}
\textsubscript{$q_{0}$} &  \qw        & \gate{X} &\gate{H}\slice{} & \meter{$Z$}\slice{} &\gate{H}& \gate{X}&  \qw      
\\
\textsubscript{$q_{1}$} & \qw &  \ctrl{-1} & \qw     & \qw & \qw &\ctrl{-1} & \qw
\end{quantikz}
\label{protocol},
\end{equation}

and the decomposition for the $M_{ZZ}$:

\begin{equation}
\begin{quantikz}
\textsubscript{$q_{0}$} &   \qw      & \gate{X}\slice{} & \qw\slice{} & \gate{X}& \qw & \qw      
\\
\textsubscript{$q_{1}$} & \qw &  \ctrl{-1}      & \meter{$Z$} &  \ctrl{-1} & \qw  & \qw
\end{quantikz}
\label{protocol2},
\end{equation}

We consider half of the gate error and decoherence and dephasing noise of the idle gates before the measurement operation and the other half afterwards. This transpilation also implies an added gate and time complexity $t_{cycle}= 4t_{M}+8t_{CX}+8t_{H}$ for $BS$ codes compared to $t_{cycle}=2t_{H}+4t_{CX}+t_{M}$ of $S$ codes.

\subsection{Analytical treatment of cavity losses in QNDM}
\label{Ap_dec}
The quantum non-demolition measurement (QNDM) can be theoretically described using input–output theory \cite{Kuhn2015}. We analyze the contributions of an atom-cavity system $\ket{\alpha}$ and weak coherent pulses interacting with the atom-cavity setup. As described in \cite{Niemietz_2021}, five populated photonic modes are considered:

\begin{enumerate}
    \item $\ket{r}$: reflection back into the cavity from the resonator,
\item $\ket{r^{0}}$: reflection back into the fiber cladding due to imperfect fiber-cavity mode matching,
\item $\ket{t}$: resonator transmission losses,
\item $\ket{m}$: absorption and scattering losses in the fiber-cavity mirrors,
\item $\ket{a}$: atomic losses.
\end{enumerate}

The coherent state amplitudes for these modes are described as follows:

\begin{align}
    r_{0_{a},1_{a}} &= \left(1 - \mu_{FC}^2 \frac{2\kappa_r}{\frac{N g^2}{i \Delta_a + \gamma} + i \Delta_c + \kappa}\right) \alpha, \\
    r^{0}_{0_{a},1_{a}} &= \sqrt{1 - \mu_{FC}^2} \mu_{FC} \frac{2\kappa_r}{\frac{N g^2}{i \Delta_a + \gamma} + i \Delta_c + \kappa} \alpha, \\
    t_{0_{a},1_{a}} &= \mu_{FC} \frac{2\sqrt{\kappa_r \kappa_t}}{\frac{N g^2}{i \Delta_a + \gamma} + i \Delta_c + \kappa} \alpha, \\
    m_{0_{a},1_{a}} &= \mu_{FC} \frac{2\sqrt{\kappa_r \kappa_m}}{\frac{N g^2}{i \Delta_a + \gamma} + i \Delta_c + \kappa} \alpha, \\
    a_{0_{a},1_{a}} &= \mu_{FC} 2g \frac{\sqrt{\kappa_r \gamma N}}{i \Delta_a + \gamma} \frac{\alpha}{\frac{N g^2}{i \Delta_a + \gamma} + i \Delta_c + \kappa}.
\end{align}

Here, $N=\{0,1\}$ is the number of atoms coupled to the cavity mode, $\gamma = 1 \times 2\pi$ MHz is the atomic dipole decay rate, and $g = 25 \times 2\pi$ MHz is the coupling strength with the qubit cavity mode. The spectral detuning between the incoming weak coherent field and the atomic transition frequency is $\Delta_a = 0.01 \times 2\pi$ MHz, and the spectral detuning between the qubit cavity mode frequency and the field is $\Delta_c = 0.01 \times 2\pi$ MHz.

The decay rates into respective modes are given as $\kappa_r = \frac{4.0}{4.3} \kappa$, $\kappa_t = 1.20 \times 2\pi$ MHz, and $\kappa_m = 1.20 \times 2\pi$ MHz, with the total field decay rate $\kappa = 27.8 \times 2\pi$ MHz. The imperfect fiber-cavity mode matching factor is $\mu_{FC} = 0.99 e^{-i 0.03}$, and the amplitude of the weak coherent pulses is $\alpha = 0.3$.

Considering the representation simplification

\begin{equation}
\ket{l_{0_{a},1_{a}}} = \ket{r_{0_{a},1_{a}}, r^{0}_{0_{a},1_{a}}, t_{0_{a},1_{a}}, m_{0_{a},1_{a}}, a_{0_{a},1_{a}}},
\end{equation}
the state after photon interaction with the atom-cavity system is expressed as

\begin{equation}
\ket{\psi^{1}_{rla}} = \frac{\ket{r_{0_{a}}, l_{0_{a}}, 0_{a}} + \ket{r_{1_{a}}, l_{1_{a}}, 1_{a}}}{\sqrt{2}}.
\label{State}
\end{equation}

To perform a non-destructive measurement experimentally, a microwave pulse is applied to the atomic part with a rotation operator phase of $\frac{\pi}{2}$. This results in the state:
\begin{dmath}
\ket{\psi^{2}_{rla}} = \frac{1}{2}(\ket{r_{0_{a}}, l_{0_{a}}, 0_{a}} + \ket{r_{0_{a}}, l_{0_{a}}, 1_{a}} \\
+ \ket{r_{1_{a}}, 
l_{1_{a}}, 0_{a}}  - \ket{r_{1_{a}}, l_{1_{a}}, 1_{a}}).
\label{State3}
\end{dmath}

Detection efficiencies can be estimated from Eq.\,(\ref{State3}). For example, the probability $P(0_{a})$ can be obtained by tracing out the reflective and photon loss modes:

\begin{align}
\rho &= \text{Tr}_{rl} \left( \ket{\psi^{2}_{rla}} \bra{\psi^{2}_{rla}} \right), \\
P(0_{a}) &= \bra{0_a} \rho \ket{0_a}.
\label{trace}
\end{align}

This estimation holds provided that dark counts are not considered.

\end{document}